\documentclass[prl,amsmath,aps,twocolumn,superscriptaddress]{revtex4}
\usepackage{graphicx,xspace}
\usepackage{float}
\usepackage{amsmath}
\usepackage{amssymb}
\usepackage[normalem]{ulem}
\input{epsf}
\begin{document}
\newcommand{\scalar}[2]{\left \langle#1\ #2\right \rangle}
\newcommand{\me}{\mathrm{e}}
\newcommand{\mi}{\mathrm{i}}
\newcommand{\dif}{\mathrm{d}}
\newcommand{\period}{\text{per}}
\newcommand{\free}{\text{fr}}
\newcommand{\mq}[2]{\uwave{#1}\marginpar{#2}} 
\include{latexcommands}

\title{Entanglement properties and momentum distributions  of hard-core anyons on a ring.}
\author{Raoul Santachiara} \email{santachi@lpt.ens.fr}
\affiliation{CNRS-Laboratoire de Physique Th\'eorique de l'Ecole Normale
  Sup\'erieure, 24 rue Lhomond, 75231 Paris, France.}  \author{Franck
  Stauffer} \email{fstauffe@uni-koeln.de} \affiliation{Institute f\"ur
  Theoretische Physik, Universit\"at zu K\"oln, Z\"ulpicherstra{\ss}e 77,
  50937 K\"oln, Germany } \author{Daniel C. Cabra}
\email{cabra@lpt1.u-strasbg.fr} \affiliation{Laboratoire de Physique
  Th\'{e}orique, Universit\'{e} Louis Pasteur, 3 Rue de l'Universit{\'e},
  67084 Strasbourg, C\'edex, France.}  \affiliation{Departamento de
  F\'{\i}sica, Universidad Nacional de La Plata, C.C.\ 67, 1900 La Plata,
  Argentina.}  \affiliation{Facultad de Ingenier\'\i a, Universidad Nacional
  de Lomas de Zamora, Cno.\ de Cintura y Juan XXIII, (1832) Lomas de Zamora,
  Argentina.}

\begin{abstract}
  We study the one-particle von Neumann entropy of a system of $N$
  hard-core anyons on a ring. The entropy is found to have a clear
  dependence on the anyonic parameter which characterizes the
  generalized fractional statistics described by the anyons. This
  confirms the entanglement is a valuable quantity to investigate
  topological properties of quantum states. We derive the
  generalization to anyonic statistics of the Lenard formula
  for the  one-particle
  density matrix of $N$ hard-core bosons in the large $N$ limit and
  extend our results by a numerical analysis of the entanglement
  entropy, providing additional insight into the problem under consideration.
\end{abstract}

\maketitle
In recent years an intense research activity has been
devoted to the study of entanglement in many-body states.
Initially, this effort has been mostly motivated by the fact that
quantum correlated many-body states, which appear in various
solid-state models, can be valuable resources for information
processing and quantum computation \cite{Bennet, Nielsen}.  The
theory of entanglement is now attracting even more attention
because of its fundamental implication for the development of new
efficient numerical methods for quantum systems
\cite{Verstraete,Vidal0, Cirac} and for the characterization of
quantum critical phases \cite{Fazio,Vidal, Fradkin}.

Generally speaking, entanglement measures nonlocal properties of
composite quantum systems and it can provide additional
information to that obtained by investigating local observables or
traditional correlation functions. In this respect entanglement
might be a sensitive probe into the topological properties of
quantum states. A particularly significant quantity is the
entanglement entropy $S_A$, which is defined in a bipartite system
$A\cup B$ and quantified as the von Neumann entropy
$S_A=-\mbox{Tr} \rho_A \ln \rho_A$ associated to the reduced
density matrix $\rho_A$ of a subsystem $A$. In two-dimensional
systems a firm connection between topological order and
entanglement entropy has been established in \cite{Kitaev, Levin},
where the entanglement entropy was defined by spatial partitioning.
Recent studies on Laughlin states
\cite{Latorre,Schoutens} have considered
 the entanglement entropy associated to particle partitioning
 \cite{Latorre,Schoutens}. Also in this case, the
entanglement entropy turns out to reveal important aspects of the
topological order in Laughlin States.

The two-dimensional case  is of particular interest due to the
existence of models whose elementary excitations exhibit
generalized fractional statistics. Anyons, the particles obeying
such statistics, play a fundamental role in the description of the
fractional quantum Hall effect \cite{Wilczek}. Although this
concept is essentially two-dimensional, anyons can also occur in
one-dimensional (1D) systems
\cite{Haldane,Zhu,Kundu,Batchelor1,Batchelor2,Batchelor3,
Girardeau}, where statistics and interactions are inextricable,
leading to strong short-range correlations.  The 1D anyonic models
have proven useful to study persistent charge and magnetic
currents in 1D mesoscopic rings \cite{Zhu}. This possibility and
their own pure theoretical interest lead us to investigate the
effects of the anyonic statistics on the entanglement entropy in
the present Letter. A discussion about quantum statistics and
entanglement in a two-fermions system was introduced in
\cite{Schliemann} and extended to the case of two bosons in
\cite{Paskauskas}. A mechanism of spin-space entanglement transfer
based on the indistinguishability of two particle was proposed in
\cite{Omar} and shown to depend on the statistics (either
fermionic or bosonic) of the particles involved. A comparison
between the entanglement of two bosons and two fermions confined
in a 1D harmonic trap has been presented recently in \cite{Sun}.

In this Letter, we consider a system of $N$ hard-core anyons on a
ring which is the direct anyonic generalization of the
Tonks-Girardeau gas. It offers a convenient framework to study
topological effects: the many body ground state is known and its
behaviour under the exchange of two particles interpolates between
bosons and fermions. We carry out an analytical and numerical
analysis of the dependence of the one-particle von Neumann entropy
on the statistical parameter which determines the symmetry of the
many body state.  We derive the large $N$ asymptotic expression of
the anyonic one-particle density matrix. This asymptotic form
generalizes the one obtained for hard-core bosons
\cite{Lenard3,Forrester1} and provides a one-parameter family of
zero temperature momentum distributions interpolating between
hard-core boson and free fermion distributions. Our results show
that particle entanglement depends in a non-trivial manner on the
statistics and as such, may prove to be relevant to the study of
topological properties of many-body quantum states.

Let us consider a 1D system of anyons confined on a ring of length
$L$ interacting each other via a repulsive $\delta-$function
potential. The model is defined by the Hamiltonian:
\begin{equation}
  H=-\sum_{i}^{N}\frac{\partial^2}{\partial x_{i}^{2}}+\gamma\sum_{1\leq i<j\leq N}\delta(x_i-x_j),
\label{model}
\end{equation}
The $N-$anyons wave function
$\Psi^{\theta}(x_1,x_2,..,x_N)$ exhibits a generalized symmetry
under the exchange of particles:
\begin{equation}
  \Psi^{\theta}(\cdots x_i,x_{i+1}\cdots)=
  -e^{i\theta \varepsilon(x_{i+1}-x_i)}
  \Psi^{\theta}(\cdots x_{i+1},x_i\cdots),
\label{anyon_exhange}
\end{equation}
where $\varepsilon(x)=0$ (or $1$) if $x>0$ ($x<0$) and $\theta$ is
the anyonic parameter, defined as in Ref. \cite{Zhu}. For
$\theta=0$ this model describes free fermions while, for
$\theta=\pi$, it reduces to the Lieb-Liniger Bose gas.

As first discussed in \cite{Kundu}, the problem of 1D anyons with
contact interactions allows for an exact Bethe ansatz solution which
shows that the Hamiltonian (\ref{model}) has the same energy
spectrum than a 1D interacting Bose gas with anyonic
statistics-dependent effective coupling in the moving frame.
Very recently, a detailed analysis of the
low-energy properties of this model has been carried out in
\cite{Batchelor1,Batchelor2,Batchelor3}.  It was shown that
the low-temperature thermodynamics of 1D anyons with a
$\delta$-function potential coincides with the one of a gas of
ideal particles obeying Haldane statistics: the interplay between
the anyonic parameter $\theta$ and the coupling constant $\gamma$
determines a continuous range of these generalized statistics.
These studies have shown that, for strong coupling, the dispersion
relations of the anyon gas remain linear in the thermodynamic
limit and the finite size corrections of the ground state energy
found a central charge $c=1$.

In the case of spatial partitioning, the subsystem $A$ being a
block of size $l$, conformal field theory results
\cite{Korepin,Calabrese} predict the entanglement entropy (block
entropy) $S_A (l)$ to scale as $S_A(l)\sim \frac{1}{3}\ln l$. The
dependence on the coupling constant and on the anyonic parameter
is expected to show up in the sub-leading terms
which, to the best of our knowledge, remain unknown. Below we will demonstrate that  
the one-particle
entanglement entropy $S_1^{\theta}(N)$ of $N-$anyons depends clearly on the
anyonic parameter. Furthermore, we will show that, for a great number N of
particles, the dependence of the entanglement entropy on the anyonic
statistics appears in the sub-leading term.

Let us consider the limit of hard-core anyons, {\it i.e.}
$\gamma\to\infty$. As recently shown in \cite{Girardeau}, the
Fermi-Bose mapping method for one dimensional hard-core bosons
\cite{Girardeau0} can be generalized to an anyon-fermion mapping
(AF).
Imposing the exclusion principle, {\it i.e.} the vanishing of the many-body wave
function when two particles occupy the same position, the AF
mapping reads \cite{Girardeau}:
\begin{equation}
  \Psi^{\theta}_{0}(x_1,\cdots,x_N)=\left[\prod_{1\leq i < j \leq N} A(x_i,x_j)
  \right]\Psi^{F}_{0}(x_1,\cdots,x_N),
  \label{AF}
\end{equation}
where $\Psi^{F}_{0}(x_1,\cdots,x_N)$ is the $N$ free fermion
ground state function and $A(x_i,x_j)= e^{i\theta
\varepsilon(x_i,x_j)}$. In the following, we restrict ourselves to
the case where $N$ is odd, which corresponds to a non-degenerate
ground state. The topological properties of the $N$-anyon wave
function are encoded in the factor $\prod_{1\leq i < j \leq N}
A(x_i,x_j)$ which gives the statistical phase $e^{i \theta P}$
resulting from the $P$ exchanges needed for the particle positions
to be brought to the ordering  $0\leq x_1\leq x_2\leq \cdots \leq
x_N \leq L$. The periodic boundary conditions
$\Psi^{\theta}(x_1,\cdots,x_i+L,\cdots,x_N)=\Psi^{\theta}(x_1,\cdots,x_i,\cdots,x_N)$
impose the anyonic parameter to be an integer multiple of
$2\pi/(N-1)$, $\theta=\frac{2\pi}{N-1} n$.

We are interested in computing the entropy
$S_1^{\theta}(N)=-\mbox{Tr}(\rho_N^{\theta} \ln \rho_N^{\theta})$,
where $\rho_N^{\theta}(x-x')$ is the one-particle reduced density
matrix:
\begin{eqnarray}
  \rho_N^{\theta}(x-x')=&& \int_{0}^{L}\ldots \int_{0}^{L}
  \prod_{i=2}^N d x_i \left[\bar{\Psi}^{\theta}(x,x_2,\cdots,x_N) \right.  \nonumber \\
  && \left.\hspace{1cm} \times\, \Psi^{\theta}(x',x_2,\cdots,x_N) \right],
  \label{def_ro}
\end{eqnarray}
normalized such that $\rho_N(0)=1.$ In momentum
space, the entanglement entropy simply reads:
\begin{equation}
S_1^{\theta}(N)=-\sum_{n=-\infty}^{\infty }c_N^{\theta}(n) \ln c_N^{\theta}(n),
\label{ent_moment}
\end{equation}
where $c^{\theta}_N(n)=1/L\int_{0}^{L} \rho_N^{\theta}(x) \cos(
2\pi/L n x)$ is the momentum occupation in the ground state. For $\theta=0$
(free fermions), one has $c^{0}_N(n)=1/N$ for $(N-1)/2\leq n \leq
(N+1)/2$ leading to the well known-result for free fermions
$S_1^{0}=\ln N$. The von
Neumann entropy actually measures the uncertainty about the quantum
state to attribute a state to the subsystem in consideration, which in
this case comes exclusively from the fact the
fermions are indistinguishable.

The mapping (\ref{AF}) between Fermi and anyon
eigenfunctions preserves all scalar products and thus the energy
spectrum and all the probability distributions
involving the norm of the wave function are identical.
Nevertheless, $c^{\theta}_N(n)$
strongly depends on $\theta$ as can be guessed from the drastic
difference between the momentum distributions
of free fermions and hard-core bosons.

\begin{figure}[ht!]
\begin{center}
\includegraphics [scale=0.6]{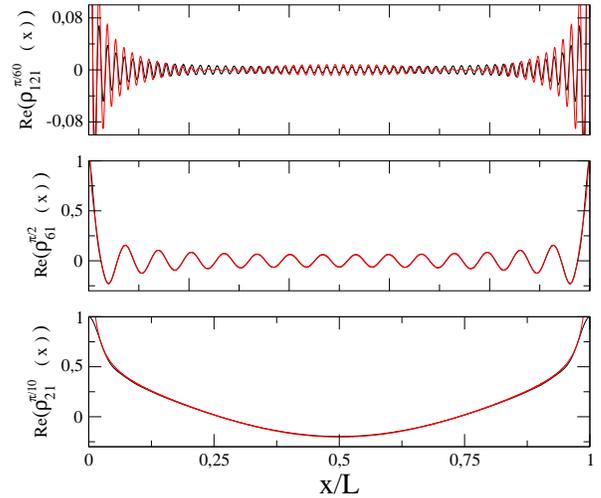}
\end{center}
\caption{(Color online) Comparison between the numerical computation
(red/dark gray curve) and asymptotic equation (black curves)
 for the one-particle
anyon density for: (a): $N=121,\theta=\pi/60$,
 (b): $N=61,\theta=\pi/2$ and (c):
$N=21,\theta=9\pi/10$. }
\label{figFvsa}
\end{figure}

Analytical expressions of $c^{\pi}(n)$ for small
($|n|\ll 1$) and large ($|n|\gg 1$) momenta have been found in the
thermodynamic limit
\cite{Lenard1,Lenard2,Sutherland,Vaidya1,Vaidya2}. These results
show a $|n|^{-1/2}$ singularity at $n=0$, reflecting the tendency
towards Bose-Einstein condensation.  The corresponding result for
finite $N$ is more cumbersome. Using the $N\gg 1$ asymptotic
result for $\rho^{\pi}(x)$ \cite{Lenard3},
\begin{equation}
\rho^{\pi}(x)\sim\rho_{\infty}N^{-\frac{1}{2}}|\sin{\pi x/L}|^{-\frac{1}{2}},
\label{as_ro_bos}
\end{equation}
with $\rho_{\infty}=G(3/2)^4/\sqrt{2}$ and $G(z)$ the Barnes
G-function \cite{G}, $c_{n}(N)$
for $N\gg n$ was shown to behave like \cite{Forrester1}:
\begin{equation}
  c_{n}^{\pi}(N)\sim \frac{\rho_{\infty}}{\sqrt{\pi}}
  \frac{\Gamma(n+1/4)}{\Gamma(n+3/4)}N^{-\frac{1}{2}},
\label{forrestercn}
\end{equation}
where $\Gamma(z)$ is the standard Gamma function.
We generalized the above results to anyonic statistics.
Representing the density matrix (\ref{def_ro})
in terms of a Toeplitz $N-1\times N-1$ determinant:
\begin{equation}
  N\rho^{\theta}_{N}(x)=\mbox{det}_{N-1}\left[\phi_{k,l}\right](x)
\label{ro_det}
\end{equation}
where
\begin{equation}
  \phi_{k,l}=\int_{0}^{2 \pi}d s\frac{2 e^{i(k-l)s}}{\pi} A(s-\frac{2\pi
x}{L})\sin\left(\frac{s}{2}-\frac{\pi x}{L}\right)\sin\left(\frac{s}{2}\right),
\label{final_form}
\end{equation}
we were able to compute the asymptotic form of $\rho^{\theta}_{N}$
using the Fisher-Hartwig conjecture \cite{Forrester2,Basor}.
For $N\gg 1$ the one-particle anyon density matrix reads:
\begin{eqnarray}
\hspace{-0.5cm}\rho^{\theta}_N(x)\sim && (2 N)^{\alpha(\theta)-1} G\left(1+\frac{\theta}{2\pi}\right)^2
 G\left(2-\frac{\theta}{2\pi}\right)^2\nonumber \\
   &&\hspace{-1cm}\times\,
 e^{i(\frac{\theta}{2\pi}-\frac{1}{2})}e^{-i N(\frac{\theta}{\pi}-1)\pi
x/L-}\left|\sin\left(\frac{\pi
x}{L}\right)\right|^{\alpha(\theta)-1}
\label{as_ro_any}
\end{eqnarray}
and $\alpha(\theta)=\frac{\theta}{\pi}(1-\frac{\theta}{2 \pi})$.
We see that Eq. (\ref{as_ro_bos})  is recovered for $\theta=\pi$.
Details of the derivation and a more complete discussion of this
result will be presented elsewhere \cite{paper2,Pasquale2}.
In the case of the generating function (\ref{final_form}) the analysis of the
behaviour of the Toeplitz determinant is subtle and the Fisher-Hartwig formula
remains a conjecture. 
The validity of 
Eq. (\ref{as_ro_any}) has thus be
compared to the numerical evaluation of the
determinant (\ref{ro_det}) for finite $N$.
For small $\theta$, the convergence to the asymptotic result is quite slow and
the formula provides a rough estimate for finite $N$, as can be seen for N=121 and
$\theta=\pi/60$ (Fig. \ref{figFvsa} (a)). The likeness increases greatly with
$\theta$ and the agreement is already perfect with $\theta=\pi/2$ for $N=61$ (Fig.
\ref{figFvsa} (b)) and for $\theta=9\pi/10$,
 close to hard-core bosons, for
N=21 (Fig. \ref{figFvsa} (c)).

\begin{figure}[ht!]
\begin{center}
\includegraphics [angle=-90,width=\columnwidth]{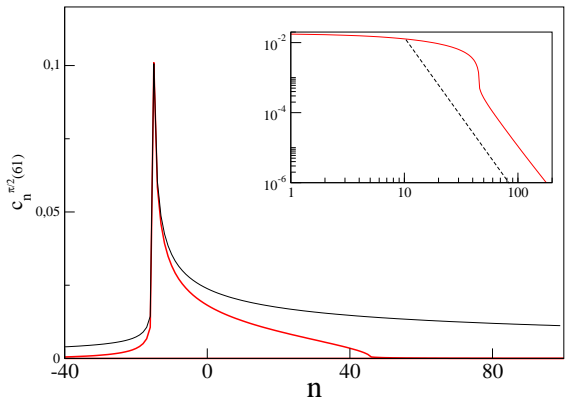}
\end{center}
\caption{$c^\theta_n(N)$ obtained from numerical (Color online) and
  Eq.(\ref{as_ro_moment}) (Black line) versus $n$ for
$\theta=\pi/2$ and $N=61$. The inset is the same plot for $n>0$ in a log-log
scale (red/dark gray curve),  while the dashed curve
is a visual guide proportional to $n^{-4}$.
}
\label{figCn}
\end{figure}

The Fourier coefficients of (\ref{as_ro_any}) can be computed
analytically \cite{Gradsthein}.  The asymptotic behaviour of
$c^{\theta}_{n}(N)$ for $N\gg n$ reads:
\begin{eqnarray}
  c_{n}^{\theta}(N)\sim &&\frac{1}{\pi}N^{\alpha(\theta)-1}
  G\left(1+\frac{\theta}{2\pi}\right)^2
  G\left(2-\frac{\theta}{2\pi}\right)^2\nonumber \\ &&
  \hspace{0.5cm}\times \Gamma\left(\alpha(\theta)\right)\sin\left(\pi (1-\alpha(\theta)\right) \nonumber \\ &&
  \hspace{1cm}\times \frac{\Gamma(n'+\theta^2/(4 \pi^2))}{\Gamma(n'+\theta/\pi-\theta^2/(4 \pi^2))}.
\label{as_ro_moment}
\end{eqnarray}
where $n'=n-\lfloor N(\theta/(2\pi) -1/2)+\theta/(4\pi)+3/4\rfloor$ with
$\lfloor x\rfloor$ being the integer part of $x$. A comparison between the
above formula and the exact results is shown in Fig. \ref{figCn}.
Note that, from Eq.(\ref{as_ro_any}), the number of particles occupying the
low-energy state ($n'=0$) scales as $N^{\alpha(\theta)-1}$, thus ruling out
the possibility of anyon-condensation predicted in the case of free anyons \cite{Mintchev}. 
For $n\gg N$, the bosonic momentum
distribution $c_n^{\pi}$ decays like $n^{-4}$
\cite{Forrester1,Anna,Olshanii}. Based on exact results for $N=5,7,9$
and on numerical computations (see Fig. \ref{figCn}), we expect this
to be true for anyonic statistics as well. Under this assumption, the terms
$c_n^{\theta}(N)$ for $n\gg N$ will not contribute significantly
to the entanglement entropy. While the main contribution to $S_1^\theta(N)$ can be extracted from
(\ref{as_ro_moment}), the numerical simulations show that, for finite $N$, the
crossover region between the asymptotic (\ref{as_ro_moment}) and power-law can not be neglected, especially near the fermionic point $\theta=0$.

\begin{figure}[ht!]
\begin{center}
\includegraphics[width=\columnwidth]{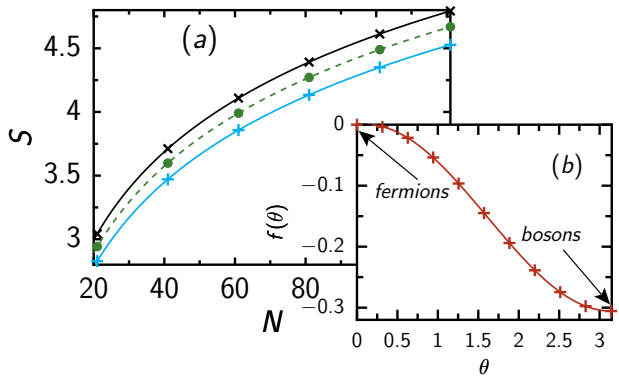}
\end{center}
\caption{(Color online) (a): Entanglement entropy as a function of N
  for $\theta=\pi$ (pluses), $\theta=\pi/2$ (crosses), $\theta=\pi/10$
  (dots) in log-linear scale, fitted according to
  Eq. (\ref{eqfit}).
  (b):   $f(\theta)$ obtained by numerical integration (pluses) and the corresponding sinus fit (plain line).}
\label{figS}
\end{figure}

We have determined numerically the value of the one-particle von
Neumann entropy $S_1^{\theta}(N)$ for different values of $N$
(Fig. \ref{figS} (a)-(b)) which can be shown to scale with $N$ in
the following way:
\begin{equation}
S_1^{\theta}(N)\approx \ln N + f(\theta) + \frac{\kappa(\theta)}{\sqrt{N}}.
\label{eqfit}
\end{equation}
The anyonic-parameter-dependent  function
$f(\theta)$ was determined numerically for system sizes for which the last term of this expansion can not be neglected. However, in the thermodynamic limit, only $f(\theta)$ will remain relevant in our discussion. The result of our numerical analysis are displayed in Fig. \ref{figS}, where panel (a) is the data as fitted by Eq. (\ref{eqfit}) for three values of the anyonic param eter and panel (b) is the resulting $f(\theta)$. This function decreases monotonically from free fermions to hard-core bosons where it respectively takes the value $f(0)=0$
and $s(\pi)\approx -0.3$, and from Fig. \ref{figS}-(b) it can be seen to be well fitted by a sinus function.

To conclude, we have investigated analytically and numerically
the one-particle von Neumann entropy and the momentum distributions
of $N$ hard-core anyons on a ring. We have determined the asymptotic
expressions for the one-particle density
matrix and for the momentum distributions. 
Numerical results show the entanglement entropy exhibits a simple
and non-trivial dependence on the anyonic parameter even for
$N\to\infty$, making it a suitable tool to study the topological
properties of many-body quantum states.

The authors thank I. Carusotto, A. Minguzzi, A. Recati and P. Scudo for
helpful discussions. In particular the authors thank P.Calabrese for important suggestions.
R.S. also thank the financial support by ANR(05-BLAN-0099-01).

\begin{thebibliography}{10}

\bibitem{Bennet}
C.~H.~Bennet and  D.~P.~DiVincenzo,
\newblock Nature  {\bf
404}, 247 (2000).

\bibitem{Nielsen}
M.~A.~Nielsen and  I.~L.~Chuang,
\newblock {\it Quantum computation and quantum information}
(Cambridge Univ. Press, Cambridge, 2000).

\bibitem{Verstraete}
F.~Verstraete,D.~Porras and J.~I.~Cirac,
\newblock   Phys. Rev. Lett. {\bf 93}, 227205 (2004)

\bibitem{Vidal0}
G.~Vidal,
\newblock Phys. Rev. Lett. {\bf 93}, 040502 (2004)

\bibitem{Cirac}
F.~Verstraete, and J.~I.~Cirac,
\newblock   Phys. Rev. B {\bf 73}, 094423 (2006).

\bibitem{Fazio}
A.~Osterloh  \textit{et al.}, 
\newblock Nature  {\bf
416}, 608 (2002).

\bibitem{Vidal}
G.~Vidal, \textit{et al.}
\newblock Phys. Rev. Lett. {\bf 90}, 227902 (2003)


\bibitem{Fradkin}
E.~Fradkin and J.~Moore
\newblock Phys. Rev. Lett. {\bf 97}, 050404 (2006).

\bibitem{Kitaev}
A.~Kitaev and J.~Preskill
\newblock Phys. Rev. Lett. {\bf 96}, 110404 (2006).

\bibitem{Levin}
M.~Levin and X.~G.~Wen
\newblock Phys. Rev. Lett. {\bf 96}, 110405 (2006).


\bibitem{Latorre}
S.~Iblisdir, J.~I.~Latorre and R.~Orus,
\newblock cond-mat/0609088.

\bibitem{Schoutens}
M.~Haque, O.~Zozulya and K.~Schoutens,
\newblock cond-mat/0609263.

\bibitem{Wilczek} F.~Wilczek, \newblock {\it Fractional Statistics and
    Anyon Superconductivity} (World Scientific, Singapore 1990).


\bibitem{Haldane}
F.~D.~Haldane,
\newblock Phys. Rev. Lett. {\bf 66}, 1529 (1991).

\bibitem{Zhu}
J.~Zhu and Z.~D.~Wang,
\newblock Phys. Rev. A {\bf 53}, 600 (1992).

\bibitem{Kundu}
A.~Kundu,
\newblock Phys. Rev. Lett. {\bf 83}, 1275 (1999).

\bibitem{Batchelor1}
M.~T.~Batchelor, X.~W.~Guan, N.~Oelkers
\newblock Phys. Rev. Lett. {\bf 96}, 210402 (2006).


\bibitem{Batchelor2}
M.~T.~Batchelor, X.~W.~Guan
\newblock Phys. Rev. B {\bf  74} 195121 (2006).

\bibitem{Batchelor3}
M.~T.~Batchelor, X.~W.~Guan
\newblock Laser Phys. Lett. {\bf 4} 77 (2007) .


\bibitem{Girardeau}
M.~D.~Girardeau,
\newblock  Phys. Rev. Lett. {\bf 97}, 210401 (2006).

\bibitem{Schliemann}
J.~Schliemann \textit{et al.} 
\newblock Phys. Rev. A {\bf 64}, 022303 (2001).

\bibitem{Paskauskas}
R.~Pauskauskas and L.~You,
\newblock Phys. Rev. A {\bf 64}, 042310 (2001).

\bibitem{Omar}
Y.~Omar,\textit{et al.} 
\newblock Phys. Rev. A {\bf 65}, 062305 (2002).

\bibitem{Sun}
B.~Sun, D.~L.~Zhou and L.~You,
\newblock Phys. Rev. A {\bf 73}, 12336 (2006).



\bibitem{Lenard3}
A.~Lenard,
\newblock  Pacific J. Math {\bf 42}, 137  (1972).

\bibitem{Forrester1}
P.~J.~Forrester, \textit{et al.} 
\newblock cond-mat/0211126.

\bibitem{Korepin}
V.~Korepin,
\newblock Phys. Rev. Lett. {\bf 92}, 096402 (2004).

\bibitem{Calabrese}
 P.~Calabrese and J.~Cardy,
\newblock J. Stat. Mech. {\bf 06}, 002 (2004).


\bibitem{Refael}
G.~Refael and J.~E.~Moore,  Phys. Rev. Lett. {\bf 93}, 260602 (2004).

\bibitem{Raullo}
 R.~Santachiara,
\newblock J. Stat. Mech. L06002 (2006).

\bibitem{Laflorencie}
 N.~Laflorencie,
\newblock Phys. Rev. B {\bf 72}, 140408R (2005).






\bibitem{Girardeau0}
M.~D.~Girardeau,
\newblock J. Math. Phys. {\bf 6}, 516 (1960).

\bibitem{Lenard1}
A.~Lenard,
\newblock  J. Math. Phys. {\bf 5}, 930  (1964).


\bibitem{Lenard2}
A.~Lenard,
\newblock  J. Math. Phys. {\bf 7}, 1268  (1966).



\bibitem{Sutherland}
B.~Sutherland,
\newblock Phys. Rev. A {\bf 4}, 2019 (1971).

\bibitem{Vaidya1}
H.~G.~Vaidya and C.~A.~Tracy,
\newblock Phys. Rev. Lett. {\bf 42}, 3 (1979).

\bibitem{Vaidya2}
H.~G.~Vaidya and C.~A.~Tracy,
\newblock Phys. Rev. Lett. {\bf 43},1540 (1979).





\bibitem{Forrester2}
P.~J.~Forrester and N.~E.~ Frankel
\newblock J. Math. Phys. {\bf 45}, 2003 (2004).

\bibitem{Basor}
E.~L.~Basor and K.~E.~Morrison,
\newblock  Linear Algebra and Its Applications {\bf 202}, 129 (1994).

\bibitem{G} 
$G(z+1)=(2 \pi)^{z/2}\mbox{exp}(-(z+(\gamma_E+1)z^2)/2)\prod_{k=1}^{\infty}(1+z/k)^k
\mbox{exp}(-z+z^2/(2k))$ where $\gamma_E$ is the Euler constant.

\bibitem{paper2}
R.~Santachiara, F.~Stauffer and D.~Cabra, unpublished.

\bibitem{Pasquale2}
We have detected an error in the derivation the sinus exponent thanks to
the comparison with the findings of Ref.\cite{Pasquale3}.

\bibitem{Pasquale3}
P~.Calabrese and M.~Mintchev,
\newblock cond-mat/0703117

\bibitem{Gradsthein}
I.~S.~Gradshtein and I.~M.~Ryzhik,
 \newblock {\it Table of integrals, series and products} (Academic Press,
 London 1980).

\bibitem{Mintchev}
A.~Liguori, M.~Mintchev and L.~Pilo,
 \newblock Nucl.Phys. B{\bf 569},577 (2000).

\bibitem{Anna}
A.~Minguzzi, P.~Vignolo and M.~P.~Tosi,
\newblock  Phys. Lett.  A {\bf  294}, 222 (2002).


\bibitem{Olshanii}
M.~Olshanii and V.~Dunjko,
\newblock Phys. Rev. Lett. {\bf 91},090401 (2003).
\end{thebibliography}
\end{document}